Long Paper

# Integrated Educational Management Tool for Adamson University

Anabella C. Doctor
Information Technology and Management Department, Adamson University
anabellacajolesdoctor@gmail.com

## Abstract

*Purpose* – This study focused on the development of a web–based integrated academic information system that can aid Adamson University faculty to become more effective and efficient in giving costless examinations, in giving student grades, in avoiding redundancy of data and efforts, and in providing accessible and reliable information about examinations and grades. The developed system automates the processes of examination and student grading.

*Method* – To achieve the goal of the study, the researcher followed the phases of software development life cycle aiming to produce high quality software output that meets or even exceeds Adamson University faculty and administrations' expectations.

*Results* – The developed system was tested in Adamson University and evaluated using the ISO/IEC 9126 software product evaluation criteria by respondents who include IT Experts and end-users with a descriptive rating of "excellent" with a mean average of 4.76 which proves that the system can be a useful tool for managing educational institutions' examination and student grading.

*Conclusion* – Integrated Educational Management Tool for Adamson University is a system that was successfully constructed using open source technology in developing web sites. The system has been successfully tested for functionality, reliability, usability, efficiency, and portability of the website with results that revealed that the developed application supports the educational institution's examination and student grading system for efficiency, reliability and accessibility.

*Recommendations* – Future studies and integration of item analysis, table of specification, and enhancement of sub-modules of the system are recommended as well as making available offline class records and exams with online auto synchronization of data processes.

*Implications* – With the utilization of a new system, Adamson University will come up with a standard institutional class record, test banking system, quality examination materials, paperless examination, class record, and test materials so that the faculty will be able to minimize time for preparing examination materials, and in checking and recording exam results. Time spent for the preparation and revision of examination materials for same subjects will also be minimized so that a smooth collaboration among fellow teachers and synchronizing of courses being taught can be achieved which will result to less deployment and implementation costing since the new system is a product of an open source technology.

# INTRODUCTION

The vast advancement of information technology is used and applied in almost all aspects of different transactions in small and large scale business organizations. Most of them have benefited from their application and implementation of an integrated system. Integrated system is a system that puts together all of an organization's systems and processes in to one complete structure that works as a single unit that can effectively and efficiently help deliver an organization's objectives. Doctor (2012) found in her study that in an academic side of the institution, integrated systems have been around for years but very few have focused on what the academic community needs to make their performance become more efficient, accessible, and reliable. Some institutions have managed to have one to several independent information systems that led to redundancies and inconsistencies of information while some still have no information systems at all. Since most schools are facing the same issues, integrated academic information system should be installed which can give advantages and answers to problems related to the system they are using.

Over the past years, diversity of assessment approaches and systems have been proposed. Rashad, Kandil, Hassan, and Zaher (2008) assert that as information technology improves, many of the educational institutions have been transformed from traditional paper–and–pencil records to computerized and web-based formats. In terms of examination, Adamson University still used and adopted the traditional ways. The instructor individually creates their examinations, and then asks for a requisition form with the chairpersons' approval, and then the faculty forwards it to the risograph department for reproduction purposes. At the moment, there is no scheme in the institutional level that is intended to preserve test questionnaires that can be retrieved later for reference purposes. In addition, most of the teachers prepare the exams, quizzes, seatworks, or assessment materials which are tasks that are being repeated over and over resulting to duplication of efforts. Assessment is one of the most crucial tasks for an institution to gauge their students' performance and competitiveness. Many instructors prepare their examinations on a buzzer beater mode, have no time to review and reflect on the prepared exams and at times, do not actually reach the objectives of the subject matters, which affect the quality of examinations, students' knowledge and performance.

In Adamson University instructors record each student's assessment and performance results, namely: grades, quizzes, major examinations score, recitation, assignments and other class standing requirements to the class record made available by the Institutional Registrar. After recording, as per requirement, the teachers submit the said class records to the department chairperson, and is then forwarded it the Registrar's Office. For a number of faculty members, they use electronic spreadsheets that automatically compute the data and manually write the same information in their class records.

Educational institutions would also only hope for the best for their students in terms of academics and there overall welfare. Students have many obstacles to overcome in order to achieve optimal academic performance. Different types of stress pose as a threat to a student's academic performance and the guidance office handles very well such situations to help students regain their focus. In the university setting, it is difficult to tell which students are suffering from these various obstacles in life that affect their academic performances. One key indicator would be the student grade as he/she progresses. Making these data systematically available to students and to guidance counselors would serve as an efficient method to monitor students' academic performance.

With the underlying situation, the Integrated Educational Management Tool (IEMT) for Adamson University is offered as a solution. It is a web-based system that integrates academic–based tasks that makes easy the formulation of examination and the facilitation of class recording.

The main objective of the study is to develop a web – based custom-made academic system that integrates and automates examination and student performance based on the standards of Adamson University. This study also specifically, aimed to: (1) design an academic integrated system as a tool that caters and integrates examination and student grading system in the academe in a more efficient, effective, and widely-accessible means with the following features, namely: (a) examination management module; (b) student grading system; and (c) report generations such as periodic grades sheets, semestral grades sheets, student grade reports, test manuscript, attendance report, and test results; (2) create a system using an open source web-based technology; (3) test the performance of the developed system based on its functions; and (4) evaluate the performance of the system using the evaluation criteria of ISO/IEC 9126 software product quality standard.

# LITERATURE REVIEW

*Information System*

Today, information system plays a vital role in an institution of higher education. Gulbahar (2008) noted that educational institutions around the world are reconsidering and restructuring the way they use information technology in order to enhance the efficiency, accessibility, reliability and accuracy of academic–based tasks. Suwardi and Permatasiri (2007) add to this by stating that numerous education institutions use and adopt information system as their means for efficiency and competitiveness of their academic activities but less of them have the means to accommodate and cater to growing institutional needs which change the organizational structure, policy, regulation, and technology itself.

Hua, Herstein, and Cassidy (2003) stress the undoubted contribution of education to the advancement and enrichment in cultural, social and economic development in all societies by endowing individuals with the means to improve their health, skills, knowledge, and capacity for productive work. Maximizing student learning in educational systems with limited resources remains one of the greatest educational challenges. This requires a constant monitoring and evaluation of the learning system in education by collecting and examining data and information used in the process of educational decision making. Information-based decision making in the management of the education system has as its goal increased access, efficiency, effectiveness, equity, and quality of education through effective systems of monitoring and evaluation, budgeting and planning, policy research and analysis. Hua et al. (2003) affirmed that Education Management Information Systems (EMIS) empowers these informed decisions to be made by providing necessary data and information by fostering an environment where the demand for information drives its use. Integrated data and information systems are at the very core of EMIS development in their support of the educational management functions throughout the education system.

*Educational Management Information System*

Monroe (2011) cited that educational management is a comprehensive effort dealing with the educational practices. It is the dynamic side of education which deals with educational institutions – right from the schools and colleges to the secretariat and is concerned with both human and material resources. The purpose of educational management is to bring pupils and teachers in a condition that will more successfully promote the realization of the objective of education. Balfour (2011) stated that the purpose of educational management is to enable the right pupils to receive the right education from the right teachers, at a cost within the means of the state, which will enable pupils to profit from their learning.

Educational Management Information System is designed to manage information about the education system and to put it to use. Hua et al. (2003) stress that EMIS is a set of formalized and integrated operational processes, procedures, and cooperative agreements by which data and information about schools and schooling, such as facilities, teachers, students, learning activities, and evaluative outputs, are regularly shared, integrated, analyzed, and disseminated for educational decision purposes that is used at every level of the educational hierarchy. The success of EMIS development can be measured and in terms of 3 factors: timely and reliable production of data and information, data integration and data sharing among departments, and effective use of data and information for educational policy decisions. Hua et al. (2003) conclude that in order to achieve the first factor for the success of EMIS, data production should be produced regularly that must meet the needs of overall educational planning and budgeting cycle, educational services, educational monitoring, evaluation and policy research and guidance in a timely fashion and must meet the needs of international collaboration and communication. The authors further explained that the timelines of meeting these needs is critically important for the success of an EMIS development and that EMIS data must come up with reliable reports from the current reality, status or trend of change of educational development of the country, district or school. The level of data reliability can be affected by almost all elements of data and information production procedures, which include the design of data collection instruments, clarity of question items, field data collection methods, educational and ethical level of respondents, design of computer database applications, data entry procedures, data aggregation methods, data integration procedures and analytical and data processing capacity. No amount of technological innovation can enhance data and information that is poor in quality from the outset. Hua et al. (2003) likewise mentioned that data integration is one of the most important parts of EMIS development, which means that data from multiple sources (payroll, achievement, and school census), multiple years, and multiple levels (student, teacher, or school level) can be linked together.

*Integration of Information System in Educational Institution*

The information system for educational institution has unique characteristics because data often change on a periodic time according to the academic process. Data will be added to the system like that on student's enrolment, while irrelevant data will be deleted after the students' graduation. At each semester, students will also change their data according to the class they choose. Suwardi and Permatasiri (2007) found that many institutions use information system to support their academic activities, but less of them have flexibility to accommodate the growth of the institution needs which change the organizational structure, policy, regulation, and the technology itself. To build an integrated information system, development study of a model is needed. Gulbahar (2008) concluded that education institutions are also seeking new strategies to support the process of technology integration, as models of technology diffusion. Suwardi and Permatasiri (2007) introduced a model of integration that caters the main activities of higher education institutions such as teaching, researching, and community servicing. From these three activities, it can be identified that the main entities of the system are students, lecturers, and non-lecturer employee. Suwardi and Permatasiri (2007) proposed the "model of process related to student activity". This model relates to three main operations of an institution including students' admission, courses, and students' graduation. Each activity involves many sub process such as student's graduation, admission planning, and admission process. Each sub processes has its sub process like determining standard of admission, room scheduling, and assignment scheduling. "Model of System Interfacing" means that after the process, data and their relationship have been identified and each process are mapped into the institutions units as the future executor of the system. This is to create the user interface of the system. The executor can be changed according to organizational changes. If the system is created first without considering the functional units of the institutions, every executor can be changed without rebuilding the system. This makes the system flexible to any organizational change.

*Information System Implementation*

Information system implementation is one of the single largest investments on higher education institutions. Financial, human resource, academic, and other information systems provide the foundation on which the business of the higher education institutions sits. Sapateiro and Goncalves (2008) cited that the higher education's business practices and processes, and the information that guides decision making in large areas of the academy, interact with and derive from these information systems. In turn, these systems and processes interact with other system's culture in ways to determine the following areas: a) how institutional resources are allocated, b) how faculty and staff interact with an institution's core business activities, c) how student needs for information and services are addressed, and d) how decision makers interact with institutional information to formulate policies and decisions. In order for the system integration to be implemented, information needed should have certain characteristics to be useful for making quality decisions.

One of the educational institutions who implemented a system integration for academic – based tasks is the Setubal Polytechnic Institute (IPS), a public institution of higher education, that was created in 1979, which composed of Presidency Services, a Welfare Social Service and five Colleges that offer a range of graduate courses in different areas such as Technology, Education, Business Administration, and Health Care. The institution has more than 6000 students, 505 teachers and 169 technical and administrative collaborators. As cited by Sapateiro and Goncalves, the idea to develop and house a new system was placed apart, because IPS did not have the sufficient skilled human resources to manage a project with all these requisites in useful time and all the systems with this kind of requisites usually take several years to become productive in a stable way. Later, a team finally found that there is only one stable, solid and tested system that meets the identified requirements to be installed in IPS–SIGARRA.

SIGARRA-Information System for the Aggregate Management of Resources and Academic registers constitute a central tool to manage superior educational establishments. It is used in all faculty members of Porto University and started to be implemented in public institutions of higher education such as IPS. SIGARRA's system architecture is an integrated Web based information system, a student's management applications (GA) and human resources management application (GRH).  Its main characteristics are information integration, web interface, modularity, and configurability.

Similarly, another study where integrated systems was developed and used for academic – based tasks is the MySchool, A Web – Based School Management Software that allows education institutions to securely manage and share information,  compile great  looking grade reports, integrate timetabling software to produce results in minutes, give parents and students online access to grades, homework, attendance and efficiently manage financial transactions that easily in touch via email, SMS or web, and produce statutory reports in a few clicks.

*Student Grading System*

Grading system is an integral part of the various tasks that need to be done by every faculty member on a regular basis.  The system involves a lot of processes such as computation of raw scores, generating subject grades, computation of GPA, determining the honor list, dean's list, student problems, top ten, seat plan, and the like. Rosas (2005) cited that grades is one of the major realities of higher education where public seems firmly convinced that students with the highest grades are best qualifies for success in life and careers. Grades are used to rank students for school admission to all courses, as merit for scholarships, points to get a job and are extolled by civic leaders. Connally (2009) stated that accurate education records do not happen by accident.  The teacher must plan and set up a system that can be easily followed and that will track grades. Interino (2004) stated in her study that student records are the school`s most important data and source of information.  It becomes an accepted fact that school information system particularly enrollment related is recognized as an asset and can be used to enhance   a school's reputation as a leading educational institution.  In the light of the aforementioned, it is incumbent upon faculty members and other academic personnel to ensure that accurate grade recording, maintenance, and other access are done.

One of the companies who produced this kind of system is the MySchool, a web-based school management software (Software Files, 2006) which makes the teachers' life easier, by simplifying the collection and production of grade reports.  The system gives teachers online access to all their resource allowing them to submit their contributions at their own time from home, school or anywhere with internet access.  It also provides tools that simplify grade report management and ensures that standard formats are followed and are error free.  Dellosa (2014) conclude that using a computer system, class record can be easily used and computation of grades can be done in a convenient way.

*Examination Management System*

Rashad et al. (2008) state that the growth of the Internet, and in particular the World Wide Web, in areas of education offers a medium that has the potential to be more responsive to students' needs.  It encourages students' greater participation in their own learning, and gives greater access to different sources of information that traditional method cannot offer. In the past decades, Rashad et al. (2008) found that variety of assessment approaches and systems have been proposed and as information technology keeps improving, many of them have been transformed from traditional paper–and–pencil to computerized and web–based format.  Rashad et al. (2008) conclude that it is necessary to build a web–based examination system for institutions/universities, as an effective solution to mass learning and evaluation of basic undergraduate education.  The web–based assessment is widely used to support students in learning and help them achieve their learning goal.  Furthermore, application of examination management systems includes assessment of the learning process itself.  Self–assessment tests are commonly used in technology enhanced environments, especially in learning management systems.  Students can use such self –assessment tests to check their acquired knowledge and get feedback about their learning process.

Examination management systems are very important for all who are involved in the education process like faculty, students, and administration staff.  For the faculty, marking the test is done automatically and instantaneously. Other benefits from the system include the following: the faculty is relieved from these time consuming duties; questions are easily edited and recycled from the question bank; while different versions of the same question can be generated for different students.  For the students, tests can be taken anytime, anywhere; questions can be attempted in a stress–less environment; and tests can be taken using a simple personal computer and the minimal requirement is just a Web browser. As for the administrator; the marks are automatically collected, analyzed, and disseminated for purposes of evaluation of teaching and learning processes.  Yuan, Zhang, and Zhan (2003) proposed a multi-layer based exam system based on Microsoft DCOM technology and found that the system is not reliable enough since it uses a specific technology and not an open source technology; the system is designed specifically for computer science students and is not designed for general students. Protic, Bojic, and Tartalja (2001) also proposed a system that provides teachers with an efficient means of generating and scoring tests with multiple choice answers and found that it is inevitable in evaluating student`s knowledge at massive examinations. Jordi, Herrera, and Josa (2001) also presented a study for securing electronic examination protocol. Using wireless technology, they propose a trade-off solution between examination security and examination flexibility. Sung, Lin, and Chen (2007) designed a prototype automatic quiz generation system for a given English text to test learner comprehension of text content and English skills and found that through the emergence of modern technologies in the field of Information Technologies, virtual learning has attained a new form.

IntelliEXAMS is the Examination Management System used by Anna University Coimbatore as their benchmark in new-age age education delivery systems and set the knowledge revolution in motion. It leverages the power of technology and facilitates the conduct of fool-proof examination, reduce-round time; enhance accuracy of evaluation and manage vast amounts of data in a distributed environment. The said Examination Management System is powered by Mindlogicx Infotech Ltd., a new generation software Product Development and Services company providing end to end solutions and services in the niche market segment of Knowledge Management and Delivery domain with the development and deployment of various products for automating the life cycle of virtual learning process. Anna University Coimbatore stated that their Examination Management System is an addition to their portfolio of solutions to provide the best to students as well as to the faculty. The Examination Management System not only places them as one of the first universities in the world having a fool proof examination. Apart from secured architecture that has been built, it offers comprehensive solutions to the university to conduct foolproof examinations without fear of any malpractice as well as help cut shorten the time and cost involved in the whole process. Rashad et al. (2008) stated that securing electronic exams are one of the most difficult challenges in a school. The relevance of the examination process for any academic institution implies that different security mechanisms must be applied in order to preserve some security properties in different examination pages. In this paper, they present a secure e-exam management system where all exams related information is presented in digital format. They proposed a cryptographic scheme that can be executed in order to achieve desired security levels at every exam production stage. Interino (2004) found that The Computerized University Entrance Examination is an alternate tool for test examination which replaces the paper-–and-pen type of examination, and enables speedy checking and scoring of the examinees answers. It provides the advantage of fast and quick generating test results to examinees and is also capable of generating statistical reports of the test scores and other related data about the examination being stored in one database.

*Evaluation Concepts*

Evaluation is a systematic determination of merit, worth, and significance of something or someone using criteria against a set of standards. It is often used to characterize and apprise subjects of interest in a wide range of enterprise, including the education and other services as cited by Capariño (2009). The author also stressed that there are professional groups who will assert the quality and firmness of the evaluation process according to the level of the topic interest. The International Standard Organization (ISO) and the International Electrical Technical Commission are the organizations who manage software standards used as gauge to measure the quality of the software developed by any programmer. ISO/IEC 9126 is the Standards for Software Engineering-Product Quality that provides an all-inclusive specification and evaluation model for the quality of software products. Zeiss et al. (2001) found that the standards ensure the 'quality of all software-intensive products' including safety critical systems are maintained.

# METHODOLOGY

*Project Design*

In any project development, designed methodology is employed to come up with effective and desirable results. The Integrated Educational Management Tool is a web-based system that integrates academic-based tasks conducted in Adamson University. The development of this system was based on an open source technology. For the meantime, this system is integrated with 2 main academic modules namely: examination management system, and student grading record system, and syllabus management and curriculum management.

To achieve this goal, current systems should be explored and evaluated first.

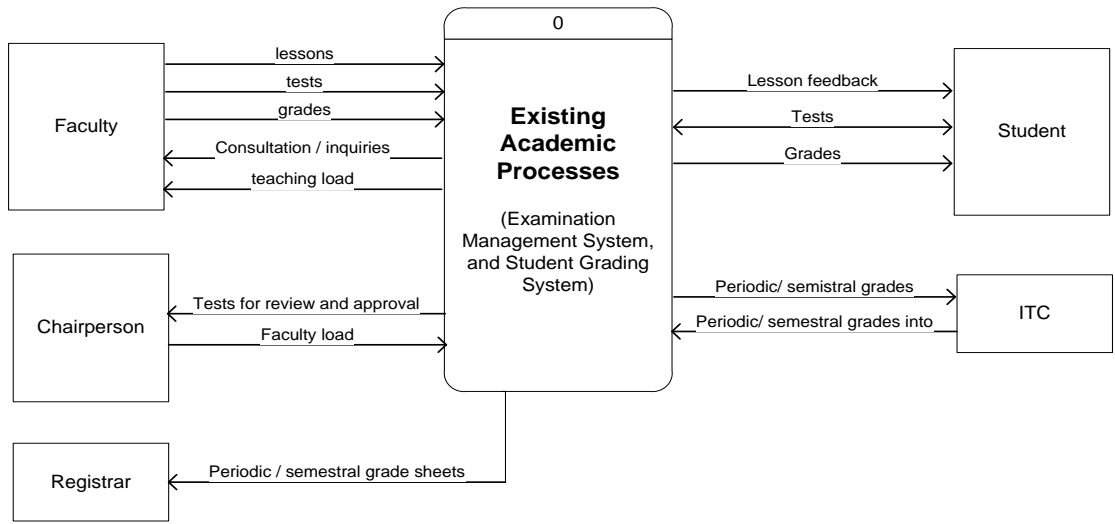

*Figure 1.* Context Data Flow Diagram of Existing Examination and Grading System of Adamson University

The existing context diagram shown in Figure 1 deals with the entities involved and the input/output flow of information into and from the academic processes as a whole. It is focused on the processes such as grading, provision of tests, loading of subjects to a faculty, admission and enlistment of students. This context data flow diagram showed the typical academic processes in an educational institution as observed by the researcher through its long experience in teaching in various institutions.

The researcher conducted the study to enhance academic processes such as student performance evaluation, recording and computation of grades. In the new system these data information from students are needed: student information, selected subject schedule, semester subject offering, new/additional subject opening, teaching assignment and teaching load.

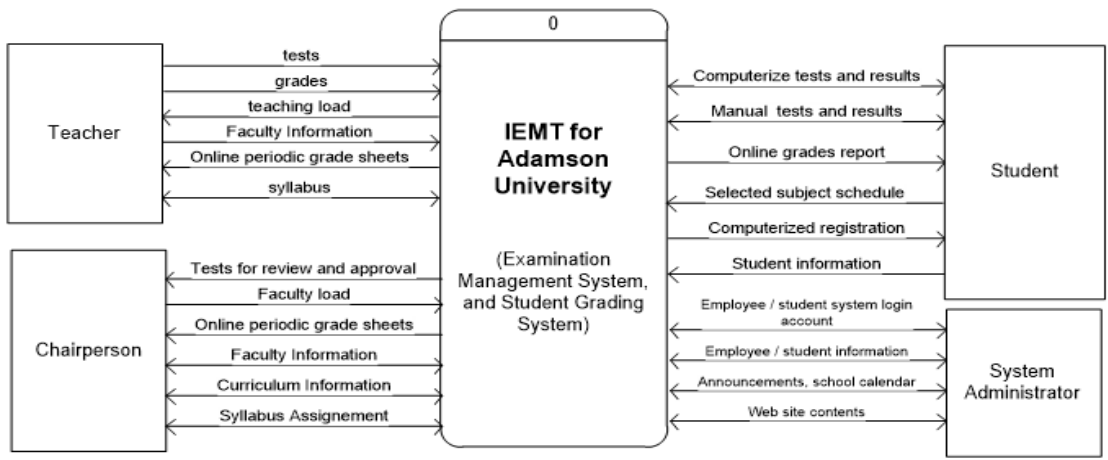

*Figure 2.* Context Flow Diagram of Integrated Educational Management Tool for Adamson University

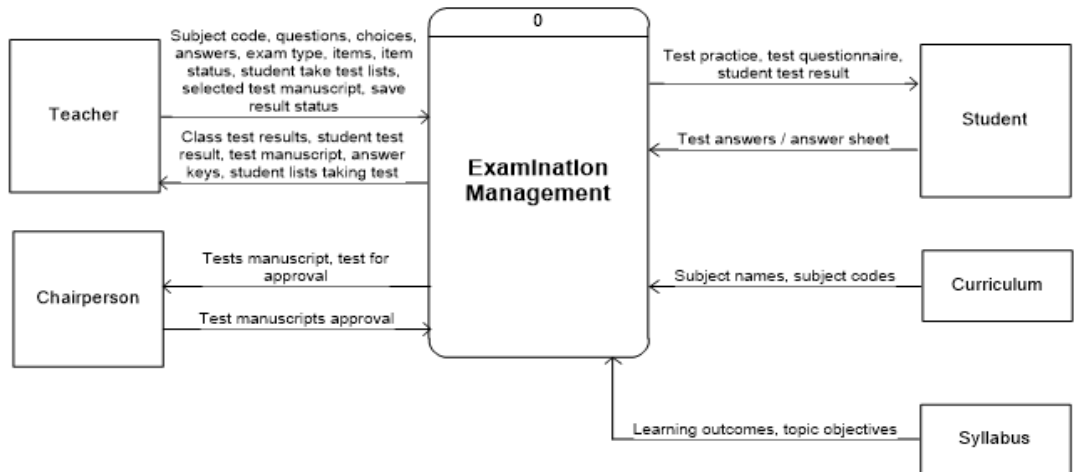

*Figure 3.a.* Context Flow Diagram of the Examination Management Module

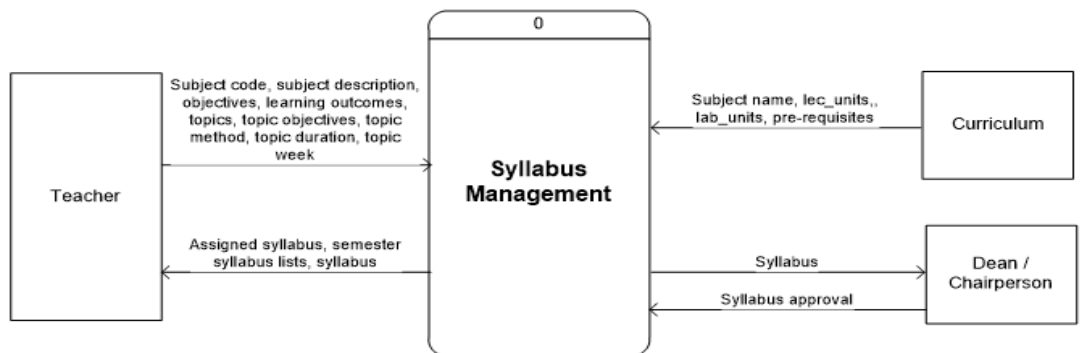

*Figure 3.b.* Context Flow Diagram of the Syllabus Management Module

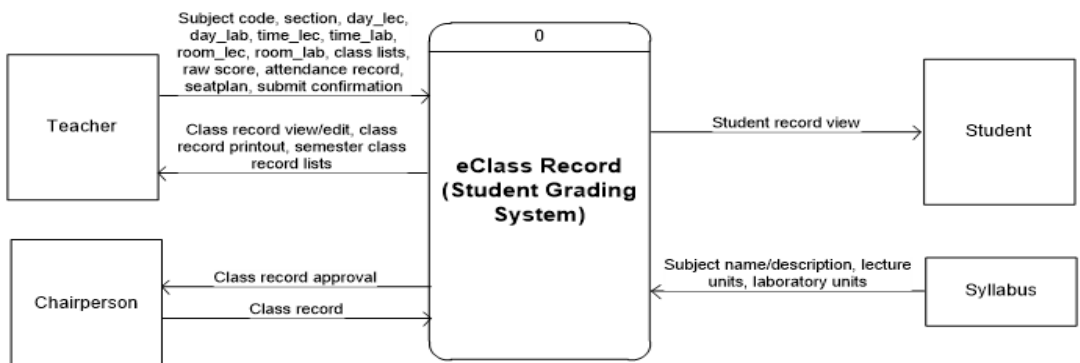

*Figure 3.c.* Context Flow Diagram of the Student Grading Management Module

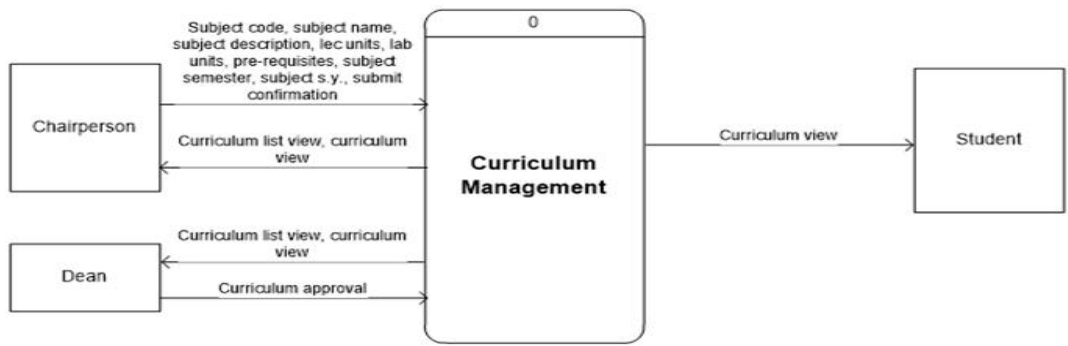

*Figure 3.d.* Context Flow Diagram of the Curriculum Management Module

Figure 2 shows the Context Flow Diagram of Integrated Educational Management Tool for Adamson University. It is the Examination Management System that is included to assist the Student Performance Evaluation. These are used for storing and retrieving test materials electronically; thus, eliminating repeated tasks for the faculty in tests/examination preparations. This system allows a computerized test for students providing significant advantages for the school; while it saves expense through a paper-less testing, thus saving paper and printing costs, and reducing time and effort for checking. The score results can be exported to the grading system for automatic recording of a student's performance. All tests are entered electronically in the system even if computerized testing is not possible. The saved questionnaires can be retrieved easily for printing, and can be used in a conventional way. The eClass Record (Student Grading System) is to be used for class record keeping and grade computation. Unlike other conventional automated grading systems, this system records all class performances including quizzes, seatworks, attendance, and other grade components; thus, providing the faculty an electronic class record. These records are extracted to automatically generate the necessary grading sheets online which students can easily access. The Curriculum Management is used for the development of courses offered by each department duly approved by both the Chairperson and the College Dean. Syllabus Management is connected to the Curriculum Management. This is used when the teacher creates/updates the course syllabi which are assigned to the faculty by the Chairperson. The database of the project system manages all the necessary data and information involved in the whole academic processes. A single repository of data provides several advantages as opposed to the multiple data store of the existing system. It eliminates data redundancy and data discrepancy. Data and information are easily made available to intended user. Since academic processes are interrelated processes, the system integration of these several processes can help the institution achieve a higher level of efficiency of data processing.

*Project Development*

In project development, careful planning should take place. The developed system features, goals, user requirements are some of the important things that should be identified in systems development. Figure 3a-Figure 3.d show the modules used by the proponent as guide to come up with an idea on how development processes of the system takes place. Module partitioning have been identified which is a very useful principle in designing a web site. The partitioned demonstrates the different tasks that have been divided and modularized in the development of the Integrated Educational Management System. The Examination Management System partition is the first plan developed by the researcher. The Syllabus Management partition was the second task, the Student Grading System is the third, and Curriculum Management is the fourth, followed by the Faculty Loading System and Content Management System of the Web site. The final phase of the development is the integration of each module that has been developed. The integration takes place in the system database.

*Operation and Testing Procedure*

The main purpose of this stage in software development before implementation and deployment takes place to ensure that the accuracy of the program meets the expected functionality, reliability, usability, efficiency and portability of the system as required. Santelices (2013) stated that Alpha and Beta testing are used to validate and verify that system meets technical requirements that guided its design and development. The test includes a step-by-step procedure on how the system is installed to the Web server by the proponent to eliminate and correct errors / faults. The procedures also include the syntax / commands used for the web designs` compatibility towards the four web browsers such as Mozilla Firefox, Safari, Google Chrome and Opera. The test conducted the different mime types used by the developer on web browsers` file handling for viewing and uploading of different file formats done as features of the system. These procedures used for testing was conducted to ensure that the desired characteristics upon running the application by the users is addressed and determined.

Testing and operation procedures are performed and shown to the evaluators during the evaluation to ensure that all features of the system functioned and performed according to the required specifications of the system. This process helped the researcher correct errors and formulated any feature enhancement of the system based from the evaluators' comments and suggestions.

*Evaluation Procedure*

There are two activities performed in the evaluation of the developed system. First, the system was presented to the evaluators-respondents which consisted of IT Professionals, system administrators, chairpersons, and deans. The functions and processes of the software are discussed during the presentation to ensure that the developed system is evaluated appropriately. Second, as suggested by the funding institution, final evaluation survey questionnaires should be distributed to at least total of 12 selected evaluators: 6 experts, 1 Department Chairperson, 1 Dean of College of Science, and 4 IT Faculty.

Suggestions and recommendations from the respondents are properly noted for the improvement of the system. The evaluation survey instrument used for this system is the criteria used in evaluating web-based system adopted from ISO 9126 Software Product Quality standard. The primary goal of the survey is to test the performance of the system from the perspective of its users. The evaluation sheet was distributed to different respondents. The questionnaire has a scale of 1 to 5. The evaluation sheet enumerates five indicators: functionality, reliability, usability, efficiency and portability.

Table 1. Likert's Scale

| Scale | Equivalent | Mean Rating Score |
|-------|-----------|-------------------|
| 5 | Excellent | 4.51 to 5.00 |
| 4 | Very Good | 3.51 to 4.50 |
| 3 | Good | 2.51 to 3.50 |
| 2 | Fair | 1.51 to 2.50 |
| 1 | Poor | 1.00 to 1.50 |

Table 1 shows the Likert's scale used in the software evaluation, its equivalent interpretation and mean score rating which is designed to capture ranges probability after averaging the scores (Ortego, 2010).

## RESULTS AND DISCUSSION

The Integrated Educational Management Tool adopted the academic standards of Adamson University. The said system has the following features:
- It has an electronic class record for summer and regular semestral class. There is an availability of class records for combined lecture and laboratory subjects and ones with pure lecture only.
- It has a class seat plan, and student attendance.
- Students are able to monitor their class performances and major examination data as shown in Figure 4 - Figure 11.

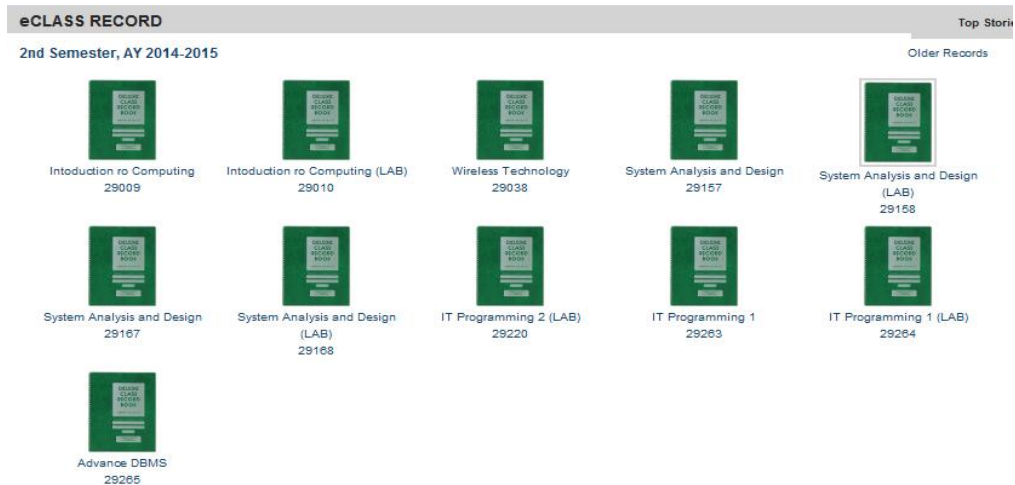

*Figure 4.* Student Grading System Management Module Interface

*Figure 5.* Class Record (Seat Plan) Interface

*Figure 6.* Class Record (Lecture and Laboratory Term Grade Sheet) Interface

*Figure 7.* Class Record (Lecture with Laboratory Semestral Grade Sheet) Interface

*Figure 8.* Class Record (Laboratory Grade Sheet) Interface

*Figure 9.* Class Record (Pure Lecture Term Grade Sheet) Interface

*Figure 10.* Class Record (Printable Term Grade Sheet) Interface

*Figure 11.* Class Record (Sending of Message to Student) Interface

- It has an examination/test management system that may be used for creating and adding questions. Editing and deleting of questions saved at the examination banks are also applicable. Different types of questions within the examinations are applicable. Examples of these are enumeration, multiple choices, true or false, and identification. Examinations can be taken by students as scheduled by their subject teacher. Examination scores can be directly linked to the electronic class record on per subject basis. Examinations created by the faculty can be shared to other faculty who are handling the same subjects. Examination result can be printed in pdf format for both the student and faculty as shown in Figure 12 - Figure 18.

*Figure 12.* Examination Module Interface

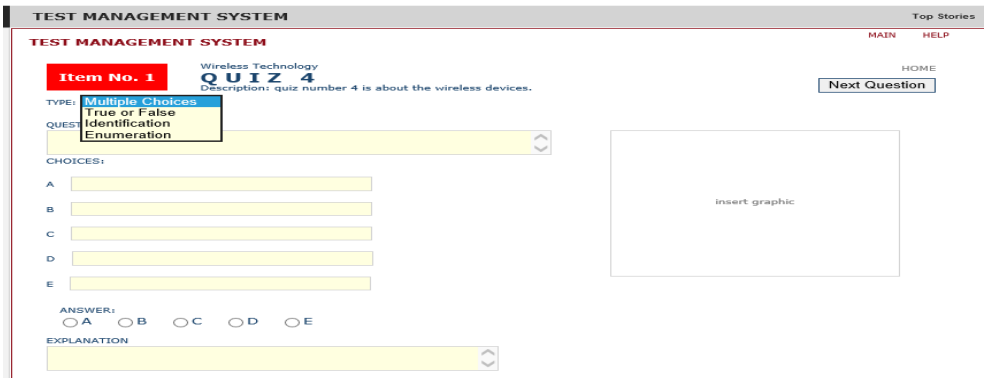

*Figure 13.* Examination Module Interface (Question Entry)

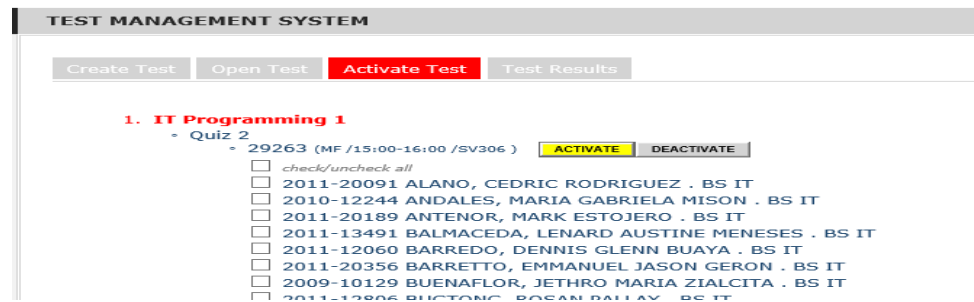

*Figure 14.* Activate Exam Interface

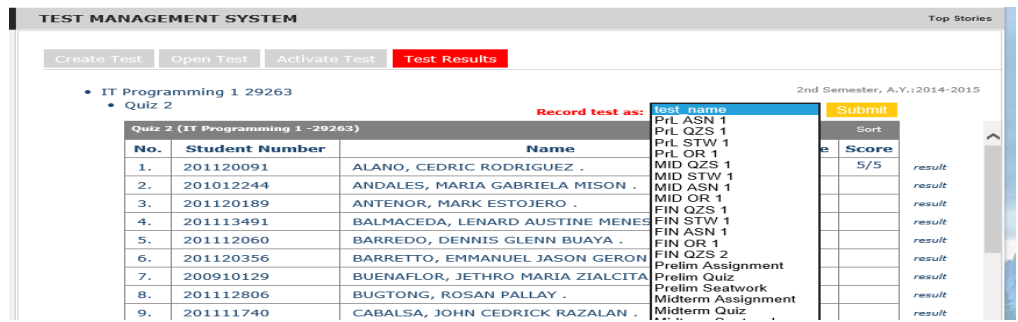

*Figure 15.* Test Result Interface

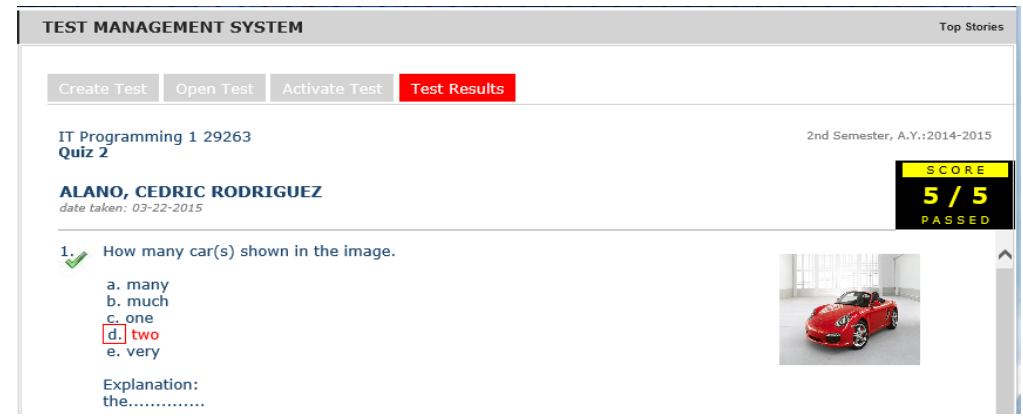

*Figure 16.* Test Result Summary Interface

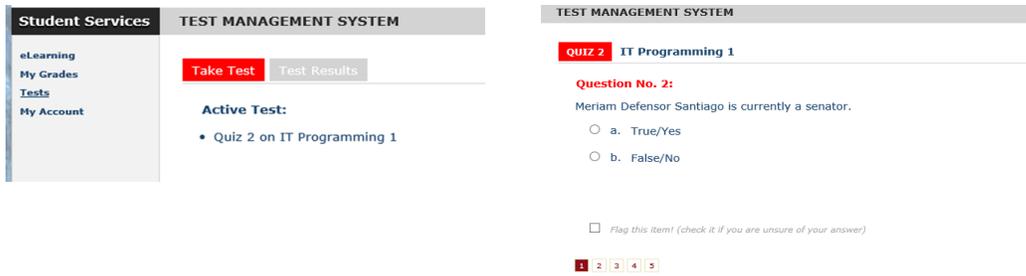

*Figure 17.* Student Active and Actual Examination Interface

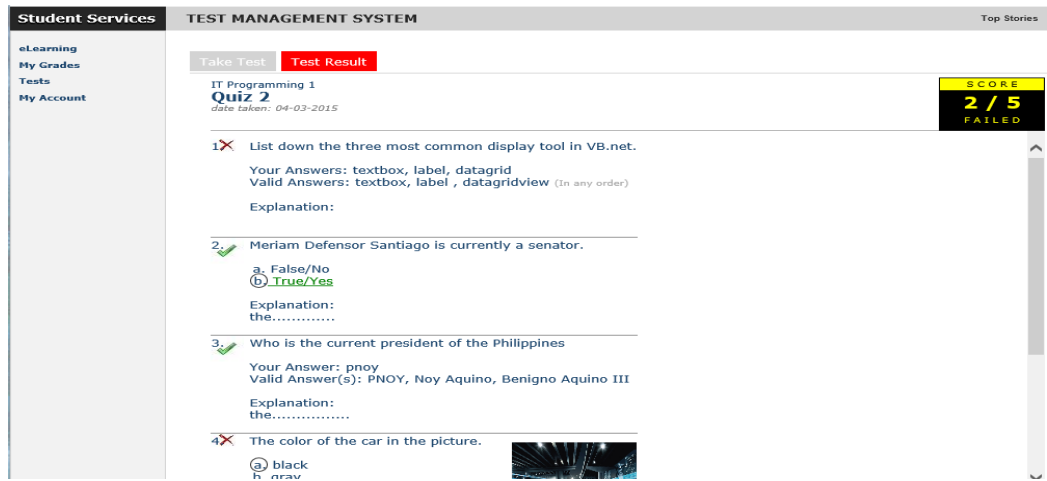

*Figure 18.* Student Examination Result Interface

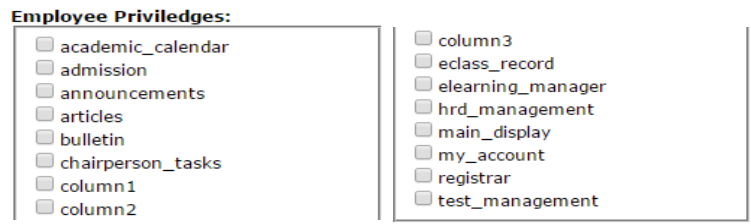

*Figure 19.* User Account Privileges Interface

- It also has a features of uploading faculty loads in excel format as part of the task of the Chairperson.
- Student Information Admission is also available. This portion of the system provides interface and processes to save students information even after the load that has been uploaded by the chairperson to his/her faculties.
- The system can register employee and its information and then register it to their respective department.
- Web – site Content Management, part of the system that manages the contents of the web site from the main page display, top stories and the article inside the main page of the system.
- Announcement Posting is also available in this system. The announcement can be viewed at the main page and the rest and web pages inside the system of IEMT.
- Curriculum and Syllabus Management is also available in this system. These features create, edit, and delete curriculum of a course offered in the institution department, and also the subject within the curriculum can be used for subject syllabus formulation by the faculty as assigned by the department chairperson.
- Events calendar that are seen at the main page of this system provides information about school events and date.
- Part of the features of the system is the creation of a user account for each employee and student with corresponding access system privilege as shown in Figure 19. This account can be used once the user opens his/her place in the system.

The developed system is composed of different users such as administrator, faculty, chairperson, and students where access to the system is based on their tasks. The administrator manages the users' account of the institution`s employees and students with appropriate privileges. He can manage the overall features of the system except the task of the chairperson and faculty. He cannot completely view the user accounts of the employees and students and has no capability to view, delete, and edit the works done by the other users of the system. The faculty is provided full capability to manage and control the operation of the examination management, student grading system module, user account editing, create as well as editing assigned course syllabus, and the sending of class and individual student messages. The faculty cannot edit E–Class Record from previous semesters and cannot view and edit the class records of other faculty including shared examinations. As for the chairperson, specific tasks and privileges are assigned to him/her and have full capabilities similar to the faculty. The chairperson manages the syllabus and the curriculum, and the uploading of faculty load. The student is given the capability to fully view their subject performances in both old and currently enrolled subjects, take examinations activated by their course teacher, view and print examination results, view curriculum and course syllabus, and send messages to classmates privately in forum manner.

For the project evaluation, 2 ways are used to evaluate the capabilities of the project which are the actual testing and the evaluation process. Actual testing involves two steps: (1) the testing during the development of the system using a standalone computer set as web server, and the testing of the system uploaded to the internet via web hosting processes, and (2) the testing is done with both local area network and uploaded to the internet via web.

Part of the actual testing is the demonstration of the project to the evaluator. The features, capabilities, and limitations of the system are explained to make sure that it is evaluated accordingly. The evaluation instrument was distributed to 12 respondents. The evaluators comprised of 6 experts, and 6 end-users within the specific department suggested by the Center of Research Department of Adamson University who funded the study. The experts are Web developers, System Administrator, and IT Technical Supports. The end-users are Department Chairperson, Dean of College of Science and other IT Faculty.

Comments, suggestions, recommendations provided by the respondents were noted for improvement purposes. The evaluation instrument for this system as shown in Table 2 is adopted from ISO 9126 software product quality criteria which is categorized into 5 categories: functionality, reliability, usability, efficiency, and portability. The criterion in the instrument provided a scale of 1-5, 5 being the highest and 1 being the lowest. The results of the evaluation resulted to the satisfaction from the IT experts and end users since they rated the project with a mark of "excellent", with an overall mean score of 4.76. Among the 5 components of the criteria, efficiency ($M = 4.84$) got the highest mean score, usability ($M = 4.83$) ranked second, functionality ($M = 4.81$) was third, reliability was fourth, and last is portability ($M = 4.65$). Refer to Table 3 for the summary of the overall mean score evaluation results.

Table 2. Evaluation Instrument Tabulated Rating of the Respondents

| Indicators | Frequency | | | | | Mean (M) | Scale | Equiv. Rating |
|---|---|---|---|---|---|---|---|---|
| | 5 | 4 | 3 | 2 | 1 | | | |
| *FUNCTIONALITY* | | | | | | | | |
| 1. Accuracy of information / data | 8 | 4 | 0 | 0 | 0 | 4.67 | 5 | E |
| 2. Availability of eClass Record | 12 | 0 | 0 | 0 | 0 | 5.00 | 5 | E |
| 3. Availability of test materials | 7 | 5 | 0 | 0 | 0 | 4.58 | 5 | E |
| 4. Capability of auto checking of examination and auto recording of test results per student and per class | 9 | 3 | 0 | 0 | 0 | 4.75 | 5 | E |
| 5. Exam questionnaire coordination with other faculty who have the same subjects | 6 | 6 | 0 | 0 | 0 | 4.50 | 4 | VG |
| 6. Can maintain information / records for future references | 6 | 6 | 0 | 0 | 0 | 4.50 | 5 | VG |
| 7. Availability of students class performance monitoring | 11 | 1 | 0 | 0 | 0 | 4.92 | 5 | E |
| 8. Capability of the system to generates exam manuscript and class records for printing purposes | 11 | 1 | 0 | 0 | 0 | 4.92 | 5 | E |
| 9. Capability of the system to generates multiple types of examination | 12 | 0 | 0 | 0 | 0 | 5.00 | 5 | E |
| 10. Can generate examination efficiently | 10 | 2 | 0 | 0 | 0 | 4.83 | 5 | E |
| 11. Can respond quickly in generating data (updating data entry, computation of grades, generation of seat plan and attendance of students) in eClass record | 11 | 1 | 0 | 0 | 0 | 4.92 | 5 | E |
| 12. Can provide security and user level access features | 8 | 4 | 0 | 0 | 0 | 4.67 | 5 | E |
| 13. Class records from other faculty cannot be seen and modified by other faculty | 11 | 1 | 0 | 0 | 0 | 4.92 | 5 | E |
| 14. Can share questionnaires from other faculty with the same subject can be viewed and activated by other faculty but cannot be modified | 11 | 1 | 0 | 0 | 0 | 4.92 | 5 | E |
| 15. Questions that are hidden are not viewable to other faculty | 11 | 1 | 0 | 0 | 0 | 4.92 | 5 | E |
| 16. Allows students to monitor their own records and does not permit to view other students' records in both examinations data and class performances | 11 | 1 | 0 | 0 | 0 | 4.92 | 5 | E |
| *Mean Average* | | | | | | **4.81** | **5** | **E** |
| *RELIABILITY* | | | | | | | | |
| 1. Procedures include no errors | 6 | 6 | 0 | 0 | 0 | 4.50 | 4 | VG |
| 2. Certain procedure level is maintained even when trouble occurs | 10 | 2 | 0 | 0 | 0 | 4.83 | 5 | E |
| 3. Normal operations are restored readily when failures occurs | 8 | 4 | 0 | 0 | 0 | 4.67 | 5 | E |
| *Mean Average* | | | | | | **4.67** | **5** | **E** |
| *USABILITY* | | | | | | | | |
| 1. The application contents can be easily understood by the users. | 10 | 2 | 0 | 0 | 0 | 4.83 | 5 | E |
| 2. The features of the application are easy to learn and are adoptable to the users. | 11 | 1 | 0 | 0 | 0 | 4.92 | 5 | E |
| 3. The application can be applied and used by the users in a fully automated process in terms of examination and class records. | 10 | 2 | 0 | 0 | 0 | 4.83 | 5 | E |
| 4. The application allots easy operation management a.Procedures are easy to process b. Reports are easy to produce | 9 | 3 | 0 | 0 | 0 | 4.75 | 5 | E |
| 5. Application designs are attractive | 10 | 2 | 0 | 0 | 0 | 4.83 | 5 | E |
| 6. The application has a user-friendly interface. | 10 | 2 | 0 | 0 | 0 | 4.83 | 5 | E |
| *Mean Average* | | | | | | 4.83 | 5 | E |
| *EFFICIENCY* | | | | | | | | |
| 1. Provides responses from procedures (in seconds) | 9 | 3 | 0 | 0 | 0 | 4.75 | 5 | E |
| 2. Allows effective resource utilization | 11 | 1 | 0 | 0 | 0 | 4.92 | 5 | E |
| *Mean Average* | | | | | | **4.84** | **5** | **E** |
| *PORTABILITY* | | | | | | | | |
| 1. The developed system can be accessed via capable mobile devices (tablets), desktop computer, laptop, and netbook. | 6 | 6 | 0 | 0 | 0 | 4.50 | 4 | VG |
| 2. The developed system can be accessed by the users via internet or intranet network connection. | 9 | 3 | 0 | 0 | 0 | 4.75 | 5 | E |
| 3. System application does not need to be installed in user's devices. | 8 | 4 | 0 | 0 | 0 | 4.67 | 5 | E |
| 4. Application is compatible with different web browsers. | 8 | 4 | 0 | 0 | 0 | 4.67 | 5 | E |
| *Mean Average* | | | | | | **4.65** | **5** | **E** |

Table 3. Summary of Overall Mean Score Evaluation from the Department of Information Technology & Management

| Indicators | Mean | Descriptive Meaning |
|---|---|---|
| A. Functionality | 4.81 | Excellent |
| B. Reliability | 4.67 | Excellent |
| C. Usability | 4.83 | Excellent |
| D. Efficiency | 4.84 | Excellent |
| E. Portability | 4.65 | Excellent |
| Overall Mean | 4.76 | Excellent |

The following is the summary of the evaluation on the five criteria shown in Table 3:

1. On functionality, the evaluators gave the application a descriptive rating of excellent with an average mean score of 4.81. This means that all features of the system are working and are functional in so far as goal of the study.

2. On reliability, the evaluators rated the system excellent. The results showed that the system conforms to the desired results, and are error free and performs well according to the specifications needed by the user. It has an average mean of 4.67.

3. On usability, the evaluators agreed that the system is adoptable and easy to learn. The system is likewise easy to understand and can be easily used by users. It has an attractive design and has a user-friendly interface. The evaluator gave an average mean of 4.83 which corresponds to excellent rating.

4. On efficiency, the evaluators gave a descriptive rating of excellent with an average mean of 4.84. This shows that the system/application conforms to the desired response of the processes and allows effective resource utilization.

5. In terms of portability, the system/application got an average of 4.65. The equivalent descriptive rating is excellent which means that the evaluators agreed that the said application or system can be accessible via internet without installing it to their computers. The system also conforms to the compatibility issues in four web browsers, such as Google Chrome, Internet Explorer, Mozilla Firefox and Opera.

## CONCLUSIONS

The following conclusions were derived based on the concerns stated in the objectives of the study and results of the evaluation conducted:

1. The Integrated Educational Management Tool for Adamson University, a web-based integrated system is designed according to the specification of Adamson University academe process whose features include:

   a. capability to assign, create, edit, delete, and open examinations for students individually or on a per class basis;

   b. capability to store exam;

   c. capability to create, edit, delete, deactivate different types of examinations such as enumeration, identification, true or false and multiple choice;

   d. interface where the examination scores of students from examination management module are automatically recorded, seat plan that are automatically arranged or rearranged, and students data at the electronic class record are viewable individually;

   e. to modify electronic class records data of the faculty easily;

   f. capability to respond faster to the examination module by auto checking current examination taken by students, providing instant exam scores and results, and automatically or manually records exam results in the class record available in the system;

   g. capability to generate randomized questionnaires for computerized examination and multi-set examination with answer keys for manual or traditional way of examination given to students;

   h. capability to retrieve previous questionnaires to serve as reviewers for practice tests for students and for faculty for future use;

   i. capability to allow students to view their own examination results summary and class performance (attendance, quizzes, major exam results, projects and others) as well as curricula, and syllabi;

   j. capable of friendly interfaces to create and edit syllabus;

k.  capability to provide online electronic class records, and exam manuscript for computerized examinations; thus, saving costs for administration;

l.  capability to show curriculum and syllabus management module;

m.  capability to manage the website where entries can be easily modified; and

n.  capability to generate periodic grade sheets, semestral grade sheets, student grade reports, test manuscripts, attendance reports, and test results.

2.  That the system was constructed under the open source technology used in developing websites for cost consideration during the implementation of the system such as: (a) Hypertext Markup Language as the lingua franca of the web; (b) Cascading Style Sheets which provides the capability to control the way in which HTML elements are laid out and displayed; (c) PHP as a server – side scripting language for Web programming; (d) Helper applications which provide supports for viewing in common graphics format; (e) JavaScript and JQuery as client – side scripting languages for Web programming; (f) MySQL as database management system software; and (g) XAMMP which provides compatibility for the PHP and MySQL run in Window base and Linux platforms;

3.  The system had been successfully tested for functionality, reliability, usability, efficiency, and portability of the website with results revealing that the developed application supports the educational institution`s academic processes for efficiency, reliability and accessibility; and

4.  The performance of the system was rated "excellent" based from the evaluation prepared and adopted from the ISO 9126 software quality standards that yielded an overall mean rating of 4.76 with excellent descriptive evaluation.

## RECOMMENDATIONS

Based on the results of the study, the following are recommended:

•  that the institution must designate a person to manage the system for information updates, and database backup when system is implemented and used;

•  the system must be uploaded to a web hosting site capable of faster response to users;

•  Examination management system scoring must be based on the given rubrics provided;

•  Future modification for improvement of the sub-module process for faculty loading, student information management be considered to increase capabilities and features of the academic integration;

•  Web page and its contents must be responsive to the resolution of the different mobile devices used presently;

•  Future studies and integration of table of specification and item analysis be shared among teachers;

•  Tracking of users activities in the system must be made available for security purposes; and

•  Availability of offline class record, and exam management automated online and offline data must be synchronized.

## IMPLICATIONS

The present integration of modern technology makes the assessment administration and student grading procedures of educational institution effective, efficient, accessible, and reliable.  With the utilization of a new system, Adamson University will come up with a standard institutional class record, test banking, high quality exam materials, paperless examination, and available class record and test materials when needed.  Faculty will spend minimal time for preparing exam materials, checking and recording exam results, preparing and revising examination materials for same subjects, and smooth collaboration among fellow teacher to synchronize subject presentations. The school administration can take advantage of the benefits derived from the system, such as less deployment and implementation costing since the new system is a product of an open source technology.

## ACKNOWLEDGEMENT

This research paper is made possible through the help and support of the personnel from the Center for Research and Development of Adamson University who funded the study and headed by Dr. Nuna E. Alamanzor, to Dr. Belinda Conde, Dr. Perlita Crusis and Prof. Marvi Aresta Bayrante who encouraged and pushed me in the completion of this project regardless of all the difficulties.  Thank you very sincerely to Dr. Carmelita Benito and to the rest of Information Technology and Management Department faculty for giving their time and effort in this study.

Above all, I thank the Almighty Father for all the wisdom, knowledge, strength, courage and faith to finish this study. To my husband for the never ending support, and to my children for their patience, prayers and sweet smiles that pushed me to work harder and continue doing this research.

# REFERENCES

Balfour, G. (2011). *Definition of education management.* Retrieved from
     http://www.preservearticles.com/2011122018637/what-is-educational-management.html.

Capariño, E. T. (2009). *Online class record system of Bulacan State University Sarmiento Campus* (Unpublished master's thesis). Technological University of the Philippines, Manila, Philippines.

Connally, B. (2009). *Information technology in the Philippines: Impact of national technology environments on business.* Retrieved from Kogod School of Business American University:
     https://es.scribd.com/document/131408758/77721260-Cblm-Tourism-1, Records 101 - 112

Dellosa, R. M. (2014). Design and evaluation of the electronic class record for LPU-Laguna International School. *Asia Pacific Journal of Multidisciplinary Research, 2*(4), 64-71.

Doctor, A. C. (2012). *Development of academic resource expert system* (Unpublished master's thesis). Technological University of the Philippines, Manila, Philippines.

Gulbahar, Y. (2008). Improving technology integration skills of prospective teachers though practice: A case study. *The Turkish Online Journal of Education Technology, 7*(4), 71-81.

Hua, H., Herstein, J., & Cassidy, T. (2003). *Educational management information system (EMIS): Integrated data and information system and their implications in educational management*. Paper presented at the Annual Conference of Comparative and International Education Society, New Orleans, L.A., USA. Retrieved from https://pdfs.semanticscholar.org/c694/a60d4d8aab9765f6ec38a7081e47e7f139f4.pdf

Interino, M. L. S. (2004). *Effectiveness of the computerized university entrance examination* (Unpublished master's thesis). Adamson University, Manila, Philippines.

Jordi, R. C., Hererra, J., & Josa, A. D. (2006). *A secure e-exam management system*. Paper presented at the *ARES '06 Proceedings of the First International Conference on Availability, Reliability and Security*, Vienna, Austria. doi: 10.1109/ARES.2006.14.

Monroe, P. (2011). *Definition of educational management.* Retrieved from
     http://www.preservearticles.com/2011122018637/what -is-educational-management.html

Ortego, M. Z. M. (2010). *Mobile phone-based Filipino-Japanese language translation system* (Unpublished master's thesis). Technological University of the Philippines, Manila, Philippines.

Protic, J., Bojic, D., & Tartalja, I. (2001). Test: Tools for evaluation of students' tests-a development experience. Paper presented at the *31st Annual Frontiers in Education Conference 2001*, Reno, NV, USA. doi: 10.1109/FIE.2001.963725

Rashad, M. Z., Kandil, M. S., Hassan, A. E., & Zaher, M. A. (2008). An Arabic web–based exam management system. *International Journal of Electrical and Computer Sciences, 10*(1), 48-45.

Rosas, M. F. (2005). *The effectiveness and adaptability of developed online grading system for a private university* (Unpublished master's thesis). Adamson University, Manila, Philippines.

Santelices, R. B. (2013). *CHEDs faculty information system in cloud computing infrastructure* (Unpublished master's thesis). Technological Institute of the Philippines, Manila, Philippines.

Sapateiro, C. M., & Goncalves, N. P. (2008). Aspects for information systems implementation: Challenges and impacts. A higher education institution experience. *Revista de Estudos Politécnicos Polytechnical Studies Review, 7*(9), 225-241.

Software Files. (2006). *MySchool web–based management*. Retrieved from http://www.myschoolmanagement.com.

Sung, L. C., Lin, Y. C., & Chen, M. C. (2007). *An automatic quiz generation system for English text*. Paper presented at the Seventh IEEE International Conference on Advanced Learning Technologies, 2007 (ICALT 2007), Niigita, Japan. doi: 10.1109/ICALT.2007.56.

Suwardi, I. S., & Permatasari, D. S. (2007). New integration model of information system on higher education institution. *Proceedings of the International Conference on Electrical Engineering and Informatics, ICEEI*

*2007*. Paper presented at the *International Conference on Electrical Engineering and Informatics*, Indonesia (pp. 654-656). School Electrical Engineering and Informatics: Bandung, Indonesia.

Yuan, Z., Zhang, L., & Zhan, G. (2003). *A novel web-based online examination system for computer science education*. Paper presented at the FIE 2003: 33rd ASEE/IEEE Frontiers in Education Conference, Westminster, CO, USA. doi: 10.1109/FIE.2003.1265999.

Zeiss, B., Vega, D., Schieferdecker, I., Neukirchen, H., & Grabowski J. (2001). *Applying the ISO 9126 quality model to test specifications exemplified for TTCN-3 test specifications*. Paper presented at the Software Engineering 2007, Hamburg, Germany. Germany: Köllen Verlag.

**Author's Biography**

Anabella C. Doctor at the moment is a fulltime permanent faculty of Adamson University. She finished two research papers and presented them in both local and international conference. Currently, she is a member of the institutional researcher and conducted innovative research funded by the Commission on Higher Education. She finished Master in Information Technology at Technological University of the Philippines, Manila Campus. She is proficient developing web-based application using open source technology and also an author of two computer books used in junior high school.